\documentclass{article}[12pt]
\usepackage{amsfonts}

\newbox\strutbox
\setbox\strutbox=\hbox{\char"B}
\def\strut{\relax\ifmmode\copy\strutbox\else\unhcopy\strutbox\fi}
\def\ialign{\everycr{}\tabskip0pt\halign}
\def\eqalign#1{\null \,\vcenter {\openup\jot \mathsurround 0pt
    \ialign{\strut \hfil$\displaystyle{##}$&$\displaystyle
      {{}##}$\hfil\crcr#1\crcr}}\,}

\def\eqalignno#1{\tabskip 0pt plus 1 fill \halign to\displaywidth{\hfil$\tabskip0pt\everycr{}\displaystyle{##}$\tabskip 0pt &$\tabskip0pt\everycr{}\displaystyle {{}##}$\hfil \tabskip 0pt plus 1 fill&\llap {$\tabskip0pt\everycr{} ##$}\tabskip 0pt \crcr #1\crcr }}

\begin{document}
\title{Quantum Amplitudes in Black-Hole Evaporation
II. Spin-0 Amplitude}
\author{A.N.St.J.Farley and P.D.D'Eath\protect\footnote{Department of Applied Mathematics and Theoretical Physics,
Centre for Mathematical Sciences,
University of Cambridge, Wilberforce Road, Cambridge CB3 0WA,
United Kingdom}}

\maketitle

\begin{abstract}
This second paper, on spin-0 amplitudes, is based on the underlying
results and methods outlined in the preceding Paper I, which describes 
the complex approach to quantum amplitudes in black-hole evaporation.  
The main result of the present paper is a computation of the quantum 
amplitude for a given slightly anisotropic configuration of a scalar field
$\phi$ on a space-like hypersurface $\Sigma_F$ at a very late time 
$T{\,}$, given also (for simplicity) that the initial data for gravity 
and the scalar field at an initial surface $\Sigma_I$ are taken to be 
spherically symmetric.  In particular, this applies to perturbations
of spherically-symmetric collapse to a black hole, starting from a 
diffuse, nearly-stationary configuration, where the bosonic part of 
the Lagrangian consists of Einstein gravity and a massless, 
minimally-coupled real scalar field $\phi{\,}$.  As described in 
Paper I, Feynman's $+i\epsilon$ approach is taken; here, this involves 
a rotation into the complex:
$T\rightarrow{\mid}T{\mid}\exp(-i\theta)$, 
with 
$0<\theta\leq\pi/2{\,}$.  A complex solution of the classical 
boundary-value problem is expected to exist, provided 
${\,}\theta >0{\,}$; although for ${\,}\theta =0{\,}$ (Lorentzian
time-separation), the classical boundary-value problem is badly posed.  
Once the amplitude is found for ${\,}\theta >0{\,},$ one can take the 
limit $\theta\rightarrow 0_{+}$ to find the Lorentzian amplitude.  
The paper also includes a discussion of adiabatic solutions of the 
scalar wave equation, and of the dimensionality of certain quantities 
which occur frequently in the project as a whole, for which these two 
initial papers establish the underlying results.
\end{abstract}

\begin{section}{Introduction}

In this treatment of spin-0 quantum amplitudes in black-hole
evaporation, a concrete calculation is given, based on the general 
procedure outlined in the previous Paper I on the complex approach to
such amplitudes [1], and in [2]; see also [3].  Consider asymptotically-flat
boundary data $(h_{ij}{\,},\phi)_I$ and $(h_{ij}{\,},\phi)_F{\,}$, 
given on two space-like boundary hypersurfaces $\Sigma_I$ and 
$\Sigma_F{\,}$, which are separated by a large Lorentzian proper-time
interval $T$, as measured near spatial infinity.  Here, 
$(h_{ij})_{I,F}=(g_{ij})_{I,F}{\;}{\,}(i,j=1,2,3)$ gives the intrinsic 
Riemannian 3-metric on the surfaces $\Sigma_I$ and $\Sigma_F{\,}$,
where $g_{\mu\nu}{\;}(\mu,\nu=0,1,2,3)$ gives the space-time 4-metric.  
As in Paper I, we assume, for the sake of definiteness, that the
bosonic part of the Lagrangian contains Einstein gravity with a 
minimally-coupled real massless scalar field $\phi{\,}$, as posed
above in the boundary data.  One would like to calculate the quantum 
amplitude corresponding to these data.  It is certainly impossible to 
compute this amplitude directly {\it via} a semi-classical expansion, 
that is, as approximately $\exp(iS_{\rm class})$, where 
$S_{\rm class}$ is the Lorentzian action of a classical solution of
the coupled Einstein/scalar field equations, subject to the boundary 
data above; typically there is no such classical solution, since the 
boundary-value problem for hyperbolic equations is badly posed [1,4,5].  
Instead, following Feynman's $+i\epsilon$ prescription [6], we rotate 
the asymptotic time-interval $T$ into the complex:
${\,}T\rightarrow{\mid}T{\mid}\exp(-i\theta)$, for 
$0<\theta\leq\pi/2{\,}$.  If the resulting complex boundary-value 
problem, up to gauge, is strongly elliptic [7], one would expect existence and 
uniqueness properties as for real elliptic equations.  In particular, 
this should give semi-classically a quantum amplitude proportional to 
$\exp(iS_{\rm class})=\exp(-I_{\rm class})$, where $I$ denotes the 
Euclidean action, provided that the Lagrangian is invariant under
local supersymmetry -- it appears that meaningful quantum amplitudes
can only be defined in that case [5,8,9]. Following Sec.I.2, one would
expect such a semi-classical amplitude to have finite loop corrections 
in a locally-supersymmetric field theory of supergravity coupled to 
supermatter, where the loop corrections will only contribute 
significantly for boundary data involving Planckian energies.  
For a pure supergravity theory, the semi-classical amplitude is 
expected to be exact [5,8,9].  Finally, to obtain the amplitude for a 
Lorentzian time-separation $T{\,}$, one takes the limit of the
amplitude as $\theta\rightarrow 0_{+}{\,}$.

The concrete calculation in this paper concerns the quantum amplitude 
corresponding to emission of scalar radiation in the black-hole 
collapse problem, as measured by non-trivial non-spherical 
perturbations of $\phi$ on the final surface $\Sigma_F{\,}$, which is 
chosen to be at an extremely late time $T{\,}$, so as to intersect all 
the outgoing radiation.  To simplify matters, for the purpose of the 
present calculation, we assume that there are no gravitons on 
$\Sigma_F{\,}$, that is, that there are no non-spherical
perturbations present in the final gravitational data $h_{ijF}{\,}$.
Equivalently, we assume here that $h_{ijF}$ describes an exactly
spherically-symmetric spatial gravitational field.  Later on, we shall 
compute the corresponding amplitudes for the opposite case, for which 
the only non-trivial perturbations in the final data are gravitational,
residing in the 3-metric $h_{ijF}{\,}$, whereas $\phi_F$ is taken to be 
exactly spherically symmetric;  this gives the amplitudes for generic 
spin-2 perturbations.  Further to simplify matters, we shall assume, 
in this paper for spin-0 and in subsequent work for spin-2, that the 
initial data $(h_{ij}{\,},\phi)_I{\,}$, representing the gravitational 
3-metric and scalar field at an early time before collapse, are
exactly spherically symmetric.  Of course, once one has the above 
amplitudes for weak spin-0 and spin-2 fluctuations separately, 
one can combine them.

In Sec.2 of the present paper, we consider the $(t,r)$-dependent
decoupled linear equations for the functions $R_{\ell m}$ appearing in
the decomposition of the scalar wave equation into angular harmonics.
Here, ${\,}\ell{\,}={\,}0,1,2,{\,}\ldots{\;}$, with $m$ subject to 
${\,}-\ell\leq{\,}m{\,}\leq{\,}\ell{\,}$, are the usual
angular-momentum quantum numbers.  These equations for
$R_{\ell m}(t,r)$ simplify considerably in the adiabatic limit, in which
$R_{\ell m}(t,r)$ oscillates much more rapidly than does the
background spherically-symmetric geometry $\gamma_{\mu\nu}{\,}$.
In this limit, the mode equation resembles the corresponding equation
for a massless scalar field in an exact Schwarzschild geometry [10,11],
except that the gravitational field and potential vary slowly with
time.  This allows one to approximate much of the classical space-time
by means of the Vaidya metric [12].  Sec.3 treats the boundary
conditions for suitable radial functions on the final surface 
$\Sigma_F{\,}$.  It also begins the process of evaluating the 
second-linearised classical action $S^{(2)}_{\rm class}$ corresponding 
to spin-0 perturbations on $\Sigma_F{\,}$, as needed in finding the 
quantum amplitude.  In Sec.4, the analytic behaviour of 
$S^{(2)}_{\rm class}$ is studied in detail, as a function of complex 
$T{\,}$, leading to an expression (2.1,4.16) for the amplitude for 
purely spin-0 perturbations on $\Sigma_F{\,}$, found by taking the 
limit $\theta\rightarrow{\,}0_{+}{\,}$, with 
$T={\,}{\mid}T{\mid}\exp(-i\theta)$.  Sec.5 is concerned, as a
practical or efficiency measure, with the question of extracting 
standard dimensionful factors from all quantities appearing in the 
calculations, so that all equations can be understood more clearly, 
reduced to relations between dimensionless quantities.  Sec.6 contains 
a short Conclusion.
\end{section}

\begin{section}{Adiabatic radial functions}

In this paper, we are mainly concerned with evaluating the scalar-field
contribution to quantum amplitudes for diffuse weak-field
gravitational/scalar data on the final surface, which is separated
from the initial surface by the
'Euclidean time-interval' $\tau{\,}$, or equivalently the 'Lorentzian
time-interval' $T{\,}$, related by 
$\tau =iT{\,}$.  Here, as in Eq.(I.2.3), we shall generally take $T$
itself to be complex, of the form ${\,}T={\mid}T{\mid}\exp(-i\theta)$, 
where ${\,}0<\theta\leq\pi/2{\,}$, in order that the classical
boundary-value problem should be well-posed.  For such amplitudes,
following Section I.5, we need to compute 
$${\rm Amplitude}{\;}{\,} 
={\;}{\,}({\rm const.})\times\exp\biggl\{i{\,}S_{\rm class}
\Bigl[(h_{ij},\phi)_{I}{\,};(h_{ij},\phi)_{F}{\,};T\Bigr]
\biggr\}{\;},\eqno(2.1)$$  
\noindent
where ${\,}h_{ij}=g_{ij}{\,}$ denotes the intrinsic 3-metric on the 
bounding surface, and [Eq.(I.5.1)]
$$S_{\rm class}[{\quad}]{\;}{\,} 
={\;}{\,}{1\over 32\pi}{\,}\biggl(\int_{\Sigma_F}-\int_{\Sigma_I}\biggr)
{\,}d^{3}x{\;}{\,}\pi^{ij}{\;}h_{ij}
+{\,}{1\over 2}\;\biggl(\int_{\Sigma_F}-\int_{\Sigma_I}\biggr)
{\,}d^{3}x{\;}{\,}\pi_{\phi}{\;}\phi{\,}-{\,}M{\,}T\eqno(2.2)$$
\noindent
gives the Lorentzian classical action, apropriate to specifying
$h_{ij}$ and $\phi$ on both bounding surfaces, each surface having the
same ADM mass $M{\,}$, where the Lorentzian time-separation between
them is $T{\,}$.  Equivalently, one may write 
${\,}{\rm Amplitude}=({\rm const.})\times{\,}\exp(-I_{\rm class})$, 
where $I_{\rm class}=-i{\,}S_{\rm class}$ is the 'Euclidean classical 
action'.

In the present context of nearly-spherical collapse to a black hole,
where the matter source is a massless scalar field $\phi{\,}$, we expand
the fields about the spherically-symmetric background solution 
$(\gamma_{\mu\nu}{\,},\Phi)$, as:
$$\eqalignno{g_{\mu\nu}{\;}{\,}&
={\;}{\,}\gamma_{\mu\nu}{\;}+{\;}h^{(1)}_{\mu\nu}{\;} 
+{\;}h^{(2)}_{\mu\nu}{\;}+{\;}\ldots{\quad},&(2.3)\cr
\phi{\;}{\,}&={\;}{\,}\Phi{\,}+{\,}\phi^{(1)}{\,}+{\,}\phi^{(2)}{\,}
+{\;}\ldots{\quad}.&(2.4)\cr}$$
\noindent
The classical action then takes the perturbative form 
$$S_{\rm class}{\;}{\,}
={\;}{\,}S^{(0)}_{\rm class}{\;}+{\;}S^{(2)}_{\rm class}{\;}
+{\;}S^{(3)}_{\rm class}{\;}+{\;}\ldots{\quad},\eqno(2.5)$$
\noindent
where $S^{(0)}_{\rm class}$ is the background action, and the
next-order correction $S^{(2)}_{\rm class}$ is of second order in
small perturbations.  Explicitly [Eq.(I.5.10)],
$$S^{(2)}_{\rm class}{\;}{\,}
={\;}{\,}{1\over
32\pi}{\,}\biggl(\int_{\Sigma_F}-\int_{\Sigma_I}\biggr)
{\;}d^{3}x{\;}{\,}\pi^{(1)ij}{\;}h^{(1)}_{ij}{\;}
+{\;}{1\over 2}{\,}\biggl(\int_{\Sigma_F}-\int_{\Sigma_I}\biggr){\;}
d^{3}x{\;}{\,}\pi^{(1)}_{\phi}{\;}\phi^{(1)}{\quad}.\eqno(2.6)$$

The quantity of physical interest is the amplitude corresponding to the
weak-field non-spherical data $\bigl(h^{(1)}_{ij}\bigr)_F$ and 
$\bigl(\phi^{(1)}\bigr)_F$ on the final surface, given here 
(approximately) by $\exp(iS^{(2)}_{\rm class})$.  For simplicity of 
exposition, we shall from now on assume that the initial data are
chosen to be exactly spherically-symmetric, given only by
$(\gamma_{ij}{\,},\Phi)_I{\,}.$  Equivalently, we now take
$$\bigl(h_{ij}^{(1)}\bigr)_{I}{\;}{\,}={\;}{\,}0{\;},
{\qquad}{\quad}\bigl(\phi^{(1)}\bigr)_{I}{\;}{\,}={\;}{\,}0.\eqno(2.7)$$   
\noindent
Then, the amplitude ${\,}\exp(iS_{\rm class})$ will only depend on the
contributions at the final surface $\Sigma_F$ in Eq.(2.6), 
[which themselves depend on
$\bigl(h^{(1)}_{ijF}{\,},\phi^{(1)}_{F}{\,};T\bigr)$].  As a practical 
matter, we could easily put $\bigl(h^{(1)}_{ij},\phi^{(1)}\bigr)_I$ 
back into the calculations that follow.  Physically, the analogous
step of 'turning back on the early-time perturbations' corresponds, in 
'particle language' rather than in the 'field language' being used in
this paper, to the inclusion of extra particles in the in-states, 
together with the original spherical collapsing matter, and asking for 
the late-time consequences.  This was first carried out by Wald [13].

In this paper, we concentrate on the scalar-field contribution to the
quantum amplitude ${\,}\exp(iS_{\rm class})$.  That is, we need to
compute
$$S^{(2)}_{\rm class,{\,}scalar}{\;}{\,} 
={\;}{\,}{1\over 2}{\,}\int_{\Sigma_F}{\,}d^{3}x{\;}{\,}
\pi^{(1)}_{\phi}{\;}\phi^{(1)}{\quad},\eqno(2.8)$$
\noindent
where the linearised perturbations 
$\bigl(h^{(1)}_{\mu\nu}{\,},\phi^{(1)}\bigr)$
obey the linearised field equations (I.3.20-22,25) about the
spherically-symmetric background $(\gamma_{\mu\nu}{\,},\Phi)$.  Here,
$\bigl(h^{(1)}_{\mu\nu}{\,},\phi^{(1)}\bigr)$ must agree with the 
prescribed final data $\bigl(h^{(1)}_{ij} \phi^{(1)}\bigr)_F$ at the 
final surface $\Sigma_F{\,}$, and be zero at the initial surface 
$\Sigma_I{\,}$.  In the Riemannian case, with a real Euclidean 
time-interval $\tau$ between $\Sigma_I$ and $\Sigma_F{\,}$, or in 
the case (I.2.3) of a complex time-interval 
${\,}T=\tau{\,}\exp(-i\theta){\,}$ between $\Sigma_I$ and 
$\Sigma_F{\,}$, where ${\,}0<\theta<\pi/2{\,}$, one expects that this 
linear boundary-value problem will be well-posed.  The other, 
gravitational, contribution
$$S^{(2)}_{\rm class,{\,}grav}{\;}{\,} 
={\;}{\,}{1\over 32\pi}{\,}\int_{\Sigma_F}d^{3}x{\;}{\,}
\pi^{(1)ij}{\;}h^{(1)}_{ij}\eqno(2.9)$$
\noindent
will be studied subsequently.

As described in Section I.4, at late times the perturbed scalar field
equation reduces to
$$\nabla^{\mu}\nabla_{\mu}{\,}\phi^{(1)}{\;}{\,} 
={\;}{\,}0{\quad},\eqno(2.10)$$  
\noindent
with respect to the spherically-symmetric background 
$\gamma_{\mu\nu}{\,}$.  Here, in contrast to the case of Secs.I.3.4, 
it is most natural to describe the gravitational field in the Lorentzian
case:
$$ds^{2}{\;}{\,} 
={\;}{\,}-{\,}e^{b(t,r)}{\,}dt^{2}{\,}+{\,}e^{a(t,r)}{\,}dr^{2}{\,}
+{\,}r^{2}{\,}\bigl(d\theta^{2}{\,}+{\,}\sin^{2}\theta{\,}d\phi^{2}\bigr)
{\;}.\eqno(2.11)$$   
Making the mode decomposition with respect to 'Lorentzian coordinates' 
${\,}(t,r,\theta,\phi){\;}$:
$$\phi^{(1)}(t,r,\theta,\phi){\;}{\,} 
={\;}{\,}{1\over r}{\;}\sum^{\infty}_{\ell=0}{\;}\sum^{\ell}_{m=-\ell}
{\;}Y_{\ell m}(\Omega){\;}R_{\ell m}(t,r){\quad},\eqno(2.12)$$
\noindent
one arrives at the $(\ell{\,},m)$ mode equation:
$$\Bigl(e^{(b-a)/2}{\,}\partial_{r}\Bigr)^{2}R_{\ell m}
-{\;}\bigl(\partial_t\bigr)^{2}R_{\ell m}
-{\,}{1\over 2}\Bigl(\partial_{t}\bigl(a-b\bigr)\Bigr)
\bigl(\partial_{t}R_{\ell m}\bigr)
-{\,}V_{\ell}(t,r){\,}R_{\ell m}{\;}{\,}={\;}{\,}0{\quad}.\eqno(2.13)$$
\noindent
Here,
$$V_{\ell}(t,r){\;}{\,} 
={\;}{\,}{{e^{b(t,r)}}\over{r^2}}{\;}\biggl({\,}\ell(\ell +1){\,}
+{\,}{{2m(t,r)}\over{r}}\biggr){\quad},\eqno(2.14)$$
\noindent
where $m(t{\,},r)$ is defined by
$$\exp\Bigl(-a(t,r)\Bigr){\;}{\,}={\;}{\,}1{\,}
-{\,}{{2m(t,r)}\over{r}}{\quad}.\eqno(2.15)$$

For high frequencies of oscillation, as measured in the
nearly-Lorentzian case, it becomes simpler to understand the solutions
of the mode equation (2.13).  Noting that the present 
$(t{\,},r)$-coordinate description is obtained by replacing $\tau$ by 
${\,}it{\,}$, consider a solution $R_{\ell m}(t{\,},r)$ of the form
$$R_{\ell m}(t,r){\;}{\,} 
\sim{\;}{\,}\exp(ikt){\;}\xi_{k\ell m}(t,r){\quad},\eqno(2.16)$$
\noindent
where ${\,}\xi_{k\ell m}(t,r)$ varies 'slowly' with respect to
$t{\,}$.  In particular, at spatial infinity, where 
$r\rightarrow\infty{\,}$, ${\,}R_{\ell m}(t,r)$ is required to reduce 
to a flat space-time separated solution in which 
$\xi_{k\ell m}(t,r)$ loses its t-dependence 
[see Eqs.(2.18,21) below].

We are studying the boundary-value problem for real scalar 
perturbations $\phi^{(1)}{\,}$, as functions of $(t,r,\theta,\phi)$,
or equivalently for real functions $R_{\ell m}(t,r)$ as in
Eqs.(2.13-15), subject to the initial condition 
${\,}\phi^{(1)}\mid_{t=0}{\,}=0$ and a final condition in which 
${\,}\phi^{(1)}\mid_{t=T}$ is prescribed, where again the final time 
${\,}T{\,}$ is of the form ${\,}T={\mid}T{\mid}\exp(-i\theta)$, 
for ${\,}0<\theta\leq\pi/2{\,}$.  For ease of visualisation, one may 
regard $\theta$ as being small and positive.  Were the propagation 
simply in flat space-time, the solution would be of the form 
$$\phi^{(1)}{\;}{\,} 
={\;}{\,}{1\over r}{\;}\sum^{\infty}_{\ell=0}{\;}
\sum^{\ell}_{m =-\ell}{\;}\int^{\infty}_{-\infty}{\,}dk{\;}{\,}a_{k\ell m}{\;}
\xi_{k\ell m}(r){\;}{{\sin(kt)}\over{\sin(kT)}}{\;} 
Y_{\ell m}(\Omega){\quad},\eqno(2.17)$$
\noindent 
where the $\{a_{k\ell m}\}$ are real coefficients and each function 
$\xi_{k\ell m}(r)$ is proportional to a spherical Bessel function
$rj_{\ell}(kr)$ [14].  In our gravitational-collapse case, 
$\xi_{k\ell m}$ becomes a function of $t$ as well as of $r$, but 
otherwise the pattern remains:
$$\phi^{(1)}{\;}{\,}
={\;}{\,}{1\over{r}}{\;}\sum^{\infty}_{\ell =0}{\;}
\sum^{\ell}_{m=-\ell}{\;}\int^{\infty}_{-\infty}{\,}dk{\;}{\,}
a_{k\ell m}{\;}\xi_{k\ell m}(t,r){\;}{{\sin(kt)}\over{\sin(kT)}}{\;} 
Y_{\ell m}(\Omega){\quad}.\eqno(2.18)$$ 
\noindent
Here, the $\{a_{k\ell m}\}$ characterise the final data:  they can be
constructed from the given ${\,}\phi^{(1)}\mid_{t=T}$ by inverting 
Eq.(2.18).  The functions ${\,}\xi_{k\ell m}(t,r)$ are defined in the 
adiabatic or large--${\mid}k{\mid}{\,}$ limit, as in the previous 
paragraph, {\it via} Eq.(2.16), where $R_{\ell m}(t,r)$ obeys the
mode equation (2.13).

More precisely, provided that $k$ is large, in the sense that the
adiabatic approximation 
$${\mid}k{\mid}{\;}{\,}
>>{\;}{\,}{1\over 2}{\;}{\mid}\dot a -\dot b{\mid}{\quad},\eqno(2.19)$$
$${\mid}k{\mid}{\;}{\,}
>>{\;}{\,}\bigl\arrowvert{{\dot\xi_{k\ell m}}\over{\xi_{k\ell m}}}
\bigl\arrowvert{\quad}, 
{\qquad}{\quad}k^{2}{\;}{\,} 
>>{\;}{\,}\bigl\arrowvert{{\ddot\xi_{k\ell m}}\over{\xi_{k\ell m}}}
\bigl\arrowvert{\quad},\eqno(2.20)$$
\noindent
holds, the mode equation reduces approximately to 
$$e^{(b-a)/2}{\;}{{\partial}\over{\partial r}}
\biggl(e^{(b-a)/2}{\;}{{\partial\xi_{k\ell m}}\over{\partial r}}\biggr){\,} 
+{\,}\bigl(k^{2}-V_{\ell}{\,}\bigr){\,}\xi_{k\ell m}{\;}{\,}
={\;}{\,}0{\;}.\eqno(2.21)$$
\noindent
Here, of course, the functions $e^{(b-a)/2}$ and $V_{\ell}$ do still
vary with the time-coordinate $t{\,}$, but only adiabatically or 
'slowly'.

As described further in [2,12,17], the geometry in the space-time
region is expected to be approximated very accurately by a Vaidya
metric [12], corresponding to a luminosity in the radiated particles
which varies slowly with time.  Such a metric can be put in the
diagonal form
$$e^{-a}{\;}{\,}={\;}{\,}1{\,}-{\,}{{2m(t,r)}\over{r}}{\quad};{\qquad}
{\quad}e^{b}{\;}{\,}={\;}{\,}\biggl({{\dot m}\over{f(m)}}\biggr)^{2}
{\,}e^{-a}{\quad},\eqno(2.22)$$
where $m(t,r)$ is a slowly-varying function, with 
${\dot m}=(\partial m/\partial t)$, and where the function $f(m)$
depends on the details of the radiation.  Then Eq.(2.19) implies that
$${\mid}k{\mid}{\;}{\,}\gg{\;}{\,}{\mid}{{\dot m}\over m}{\mid}{\quad},
\eqno(2.23)$$
\noindent
provided that $2m(t,r)<r<4m(t,r){\,}$.  In this case, the rate of change
of the metric with time is slow compared to the typical frequencies of 
the radiation; further, the time-variation scale of the background
space-time $\gamma_{\mu\nu}(x)$ is much greater than the period of the
waves.  With frequencies of magnitudes ${\mid}k{\mid}{\,}\sim m^{-1}$ 
dominating the radiation, and with ${\mid}\dot m{\mid}$ of order 
$m^{-2}$ [15], the adiabatic approximation is equivalent to 
$m^{2}\gg 1{\,}$, which corresponds to the semi-classical
approximation.  If, as expected [15], $m^3$ is a measure of the time 
taken by the hole to evaporate, then $r<4m\ll m^{3}{\,}$, provided
that $m^{2}\gg 1{\,}.$  Thus, in the large$-k$ approximation used in 
deriving Eq.(2.21), it is valid at lowest order to neglect 
time-derivatives of the background metric, out to radii small compared 
with the evaporation time of the hole and with the time since the hole 
was formed.

It is natural to define a generalisation $r*$ of the standard
Regge-Wheeler coordinate $r^{*}_{s}$ for the Schwarzschild geometry [10,16],
according to 
$${{\partial}\over{\partial r^{*}}}{\;}{\,}
={\;}{\,}e^{(b-a)/2}{\;}{{\partial}\over{\partial r}}{\quad}.\eqno(2.24)$$
\noindent
Under the above conditions, the time-dependence of $r^{*}(t,r)$ is
negligibly small, and one has ${\;}r^{*}{\,}\sim{\,}r^{*}_{s}{\;}$ 
for large $r{\,}$, where
$$r^{*}_{s}{\;}{\,} 
={\;}{\,}r{\,}+{\,}2M\ln\Bigl(\bigl(r/2M\bigr)-1\Bigr)\eqno(2.25)$$
\noindent
is the Regge-Wheeler coordinate, expressed in terms of the
Schwarzschild radial coordinate $r{\,}$.  In terms of the variable 
$r^*{\,}$, the approximate (adiabatic) mode equation (2.21) reads
$${{\partial^{2}\xi_{k\ell m}}\over{\partial r^{*2}}}{\;}
+{\;}\bigl(k^{2}{\,}-{\,}V_{\ell}\bigr){\;}\xi_{k\ell m}{\;}{\,} 
={\;}{\,}0{\quad}.\eqno(2.26)$$

\end{section}

\begin{section}{Boundary conditions}

We consider here, for definiteness, a set of suitable radial functions
$\{\xi_{k\ell m}(r)\}$ on the final surface $\Sigma_F{\,}$.  As above,
since the mode equation (2.13) does not depend on the quantum number
$m{\,}$, we may choose ${\,}\xi_{k\ell m}(r){\;}={\;}\xi_{k\ell}(r){\;},$
independently of $m$.

We seek a complete set, such that any smooth perturbation field
$\phi^{(1)}(T,r,\theta,\phi)$ as in Eqs.(2.12,13), restricted to the 
final surface $\{t = T\}$, of rapid decay near spatial infinity, can
be expanded in terms of the $\xi_{k\ell m}(r)$.  The 'left' boundary 
condition on the radial functions $\{\xi_{k\ell}(r)\}$ is that of 
regularity at the origin $\{r = 0\}$:
$$\xi_{k\ell}(0){\;}{\,}={\;}{\,}0{\quad}.\eqno(3.1)$$
\noindent
The solution to the radial equation which is regular near the origin is:
$$\xi_{k\ell}(r){\;}{\,}={\;}{\,}r{\,}\phi_{k\ell}(r){\;}{\,}
\propto{\;}{\,}r{\,}j_{\ell}(kr){\;}{\,}
\propto{\;}{\,}\biggl\{\Bigl({\rm (const.)}\times{\,}(kr)^{\ell +1}\Bigr) 
+{\,}O\bigl((kr)^{\ell +3}\bigr)\biggr\}{\quad},\eqno(3.2)$$ 
\noindent
where, again, the $j_\ell$ are spherical Bessel functions [14], and we have
assumed that, for small $r{\,}$, $\;m(r)\sim r^3{\,}$, and have neglected
$O(r^2)$ terms.  These radial functions are purely real, for real $k$
and $r{\,}$.  For $k$ purely real and positive, the radial functions
describe standing waves, which, for mode time-dependence $e^{\pm ikt}$,
have equal amounts of 'ingoing' and 'outgoing' radiation.

For the 'right' boundary condition, note that the potential
$V_{\ell}(r)$ vanishes sufficiently rapidly as $r\rightarrow\infty$
that a real solution to Eq.(2.26) behaves according to 
$$\xi_{k\ell}(r){\;}{\,} 
\sim{\;}{\,}\Bigl(z_{k\ell}{\;}\exp(ikr^{*}_{s}){\;}
+{\;}z^{*}_{k\ell}{\;}\exp(-ikr^{*}_{s})\Bigr)\eqno(3.3)$$
\noindent
as $r\rightarrow\infty{\,}$.  Here the $z_{k\ell}$ are certain
dimensionless complex coefficients, which can be determined {\it via}
the differential equation by using the regularity at $\{r = 0\}$.  The
(approximately) conserved Wronskian for Eq.(2.26), together with
Eq.(3.3), and the property 
$$\lim_{r^{*}_{s}\rightarrow\infty} 
{{e^{i(k-k^{\prime})r^{*}_{s}}\over{(k-k^{\prime})}}}{\;}{\,} 
={\;}{\,}i\pi{\;}\delta(k-k'){\quad},\eqno(3.4)$$
\noindent
give the normalisation condition, for 
${\,}-\infty{\,}<{\,}k,k^{\prime}{\,}<{\,}\infty$
and $R_{\infty}\rightarrow\infty{\,}$: 
$$\int^{R_\infty}_{0}dr{\;}{\,}e^{(a-b)/2}{\;}\xi_{k\ell}(r){\;}
\xi^{*}_{k'\ell}(r)\Bigl\arrowvert_{\Sigma_F}{\;}{\,}
={\;}{\,}2\pi{\;}{\mid}z_{k\ell}{\mid}^{2}{\;}
\Bigl(\delta(k-k^{\prime})+\delta(k+k^{\prime})\Bigr){\quad}.\eqno(3.5)$$
\noindent
This normalisation is only possible in an adiabatic approximation.
Note that the radial functions $\{\xi_{k\ell}\}$ form a complete set
only for $k>0{\,}$, as a result of our boundary conditions.

The above result makes it possible to evaluate the perturbative
massless-scalar contribution to the total classical Lorentzian action
${\,}S_{\rm class}=S^{(0)}_{\rm class}+S^{(2)}_{\rm class}+\ldots{\,}$ of
Eq.(2.5), with $S^{(2)}_{\rm class}$ given by Eq.(2.6).  This 
contribution, namely $S^{(2)}_{\rm class,{\,}scalar}$ of Eq.(2.8), is then
given in the notation of Eq.(2.12) by 
$$S^{(2)}_{\rm class}\bigl[\phi^{(1)}{\,};{\,}T\bigr]{\;}{\,}
={\;}{\,}{1\over 2}{\,}\sum^{\infty}_{\ell =0}{\;}\sum^{\ell}_{m=-\ell}{\;}
\int^{R_{\infty}}_{0}dr{\;}{\,}e^{(a-b)/2}{\;}R_{\ell m}{\;}
\bigl(\partial_{t}R^{*}_{\ell m}\bigr)\Bigl\arrowvert_{T}{\quad},\eqno(3.6)$$
since
$$\int d\Omega{\;}{\,}Y_{\ell m}{\;}Y^{*}_{\ell^{\prime}m^{\prime}}{\;}{\,} 
={\;}{\,}\delta_{\ell\ell^{\prime}}{\;}\delta_{mm^{\prime}}{\quad}.
\eqno(3.7)$$
\noindent
Within the adiabatic approximation above, and using Eq.(3.5), this
gives the frequency-space form of the classical action:
$$S^{(2)}_{\rm class}\Bigl[\{a_{k\ell m}\};T\Bigr]{\;}{\,}
={\;}{\,}\pi{\;}\sum^{\infty}_{\ell =0}{\;}\sum^{\ell}_{m=-\ell}{\;}
\int^{\infty}_{0}dk{\;}{\,}k{\;}{\mid}z_{k\ell}{\mid}^{2}{\;}
{\mid}a_{k\ell m}+{\,}a_{-k\ell m}{\mid}^{2}{\;}\cot(kT){\quad},
\eqno(3.8)$$
\noindent
in terms of the final data $\{a_{k\ell m}\}$.

From a mathematical point of view, one would expect to work only with
the set of square-integrable scalar wave-functions on the final
boundary $\Sigma_F{\,}$, that is, the set 
$L^{2}(\Bbb R^{3},{\,}dr{\,}e^{(a-b)/2})$.  To express this, define
$$\psi_{\ell m}(r){\;}{\,} 
={\;}{\,}r{\,}\int d\Omega{\;}{\,}Y_{\ell m}(\Omega){\;}
\phi^{(1)}(t,r,\Omega)\Bigl\arrowvert_{t = T}{\quad}.\eqno(3.9)$$
\noindent
Then the square-integrability condition reads
$${{1}\over{2\pi}}{\;}\sum_{\ell m}{\;}\int^{R_\infty}_{0}dr{\;}{\,}
e^{(a-b)/2}{\;}{\mid}\psi_{\ell m}(r){\mid}^{2}{\;}{\,}<{\;}{\,}\infty{\quad},
\eqno(3.10)$$
\noindent
or, equivalently,
$$\sum_{\ell m}{\;}\int^{\infty}_{-\infty}dk{\;}{\,}
{\mid}z_{k\ell}{\mid}^{2}{\;}
{\mid}a_{k\ell m}+a_{-k\ell m}{\mid}^{2}{\;}{\,}<{\;}{\,}\infty{\quad}.
\eqno(3.11)$$

The left-hand sides of Eqs.(3.10,11) are in fact equal. This arises from
the completeness property
$$e^{(a-b)/2}{\;}\int^{\infty}_{-\infty}{\,}dk{\;}{\,}
{{\xi_{k\ell}(r){\;}\xi_{k\ell}(r^{\prime})}
\over{{\mid}z_{k\ell}{\mid}^{2}}}{\;}{\,} 
={\;}{\,}4\pi{\;}\delta(r-r^{\prime})\eqno(3.12)$$
\noindent
and the inverse of Eq.(2.18):
$$a_{k\ell m}{\,}+{\,}a_{-k\ell m}{\;}{\,}
={\;}{\,}{{1}\over{2\pi{\mid}z_{k\ell}{\mid}^{2}}}{\;}
\int^{R_{\infty}}_{0}{\,}dr{\;}{\,}e^{(a-b)/2}{\;}
\xi_{k\ell}(r){\;}\psi_{\ell m}(r){\quad}.\eqno(3.13)$$

From a physical point of view, one expects also that taking scalar
boundary data which are not square-integrable will lead to various
undesirable properties, such as infinite total energy of the system,
or an infinite or ill-defined action.
\end{section}

\begin{section}{Analytic continuation}

The perturbative classical scalar action $S^{(2)}_{\rm class}$ of
Eq.(3.8) was derived subject to the adiabatic approximation, and also
to the requirement that the time-interval $T$ between the initial and
final surfaces, measured at spatial infinity, is complex, of the form 
$T={\mid}T{\mid}\exp(-i\theta)$, provided that
$0<\theta\leq\pi/2{\,}$.  In this case, the term ${\,}k\cot(kT)$ in
the integrand of Eq.(3.8) remains bounded for ${\,}0<k<\infty{\,}$,
and one expects to obtain a finite complex-valued action 
$S^{(2)}_{\rm class}\Bigl[\{a_{k\ell m}\};T\Bigr]$, given
square-integrable data ${\,}\phi^{(1)}$ on the final surface 
$\Sigma_F{\,}$.  Further, the dependence of the complex function
$S^{(2)}_{\rm class}\Bigl[\{a_{k\ell m}\};T\Bigr]$ on the complex 
variable ${\,}T$ is expected to be complex-analytic in this domain 
$(0<\theta\leq\pi/2)$, and, following Feynman [6], ordinary
Lorentzian-signature quantum amplitudes should be given by the
limiting behaviour of ${\,}\exp\bigl(iS^{(2)}_{\rm class}\bigr)$ as 
${\,}\theta\rightarrow 0_{+}{\;}$.

If, on the other hand, one restricts attention to the exactly
Lorentzian-signature case $(\theta =0)$, then the integral in Eq.(3.8)
will typically diverge, due to the simple poles on the real-frequency
axis at
$$k{\;}{\,}={\;}{\,}k_{n}{\;}{\,}={\;}{\,}{n\pi\over T}{\quad},\eqno(4.1)$$
$(n=1,2,\ldots{\;}){\,}$.

Clearly, in order to be able to treat this classical
Einstein/massless-scalar boundary-value problem in a way which is
analytically sensible, one has to allow $T$ to be displaced into the
complex as above, whether by a small or a large angle $\theta$ of
rotation.  The spherically-symmetric 'background' 4-geometry
$\gamma_{\mu\nu}$ and scalar field $\Phi$ will typically be complex.
One might hope that the general (asymmetric) boundary-value problem of
this type would admit a unified mathematical description, based on the
(conjectured) property that the boundary-value problem is strongly 
elliptic [7], up to gauge.

Given that ${\,}T={\mid}T{\mid}\exp(-i\delta){\,}$ is slightly
complex, with ${\,}0<\delta\ll 1{\,}$, consider an integral such as 
Eq.(3.8) for $S^{(2)}_{\rm class}\Bigl[\{a_{k\ell m}\};T\Bigr]{\,}$.   
Write this as 
$$J{\;}{\,}
={\;}{\,}\sum_{\ell m}{\;}\int^{\infty}_{0}{\,}dk{\;}{\,}f_{\ell m}(k){\;}
\cot(kT){\quad},\eqno(4.2)$$ 
\noindent
where 
$$f_{\ell m}(k){\;}{\,} 
={\;}{\,}\pi k{\;}{\mid}z_{k\ell}{\mid}^{2}{\;}{\,}
{\mid}a_{k\ell m}{\,}+{\,}a_{-k\ell m}{\mid}^{2}{\quad}.\eqno(4.3)$$   
\noindent
There are infinitely many simple poles of the integrand at 
$\;k=k_n{\;}{\;}{\,}(n=1,2,\ldots{\;})$, just above the positive 
real $k$-axis.  We then deform the original contour $C$ along the
positive real $k$-axis into three parts, 
${\,}C_{\epsilon}{\,},{\,}C_{R}{\,}$ and ${\,}C_{\alpha}{\,}$, where 
${\,}0<\alpha\ll 1{\,}$.  The contour $C_\epsilon$ lies in the lower 
half-plane, half-encircling each of the simple poles near the positive real
$k$-axis, with radius $\epsilon{\,}$.  The curve $C_{R}{\,}$, also in the
lower half-plane, is an arc of a circle ${\,}{\mid}k{\mid}=R$ of large
radius.  The curve $C_\alpha$ is part of the radial line 
${\,}\arg(k)=-\alpha{\,}$.  We write
$$\eqalign{J{\;}{\,}&
={\;}{\,}\sum_{\ell m}{\;}\int_{{C_\alpha}{\;}+{\;}C_{R}{\;}
-{\;}C_{\epsilon}}{\;}dk{\;}{\,}f_{\ell m}(k){\;}\cot(kT)\cr
&={\;}{\,}J_{\alpha}{\;}+{\;}J_{R}{\;}+{\;}J_{\epsilon}{\quad}.\cr}\eqno(4.4)$$

Starting with the integral $J_R{\,}$, one finds 
$${\mid}J_{R}{\mid}{\;}{\,} 
\leq{\;}{\,}\sum_{\ell m}{\;}\int^{\alpha}_{0}{\;}d\theta{\;}{\,} 
R{\;}{\mid}f_{\ell m}(R,\theta){\mid}{\;}
\coth\bigl({\mid}T{\mid}R\sin\theta\bigr){\quad},\eqno(4.5)$$
\noindent
where ${\,}k{\,}={\,}R{\,}e^{-i\theta}{\,}$ on $C_{R}{\,}$, and we
have used 
${\,}{\mid}\cot(kT){\mid}{\,}
\leq{\,}\coth\bigl({\mid}T{\mid}R\sin\theta\bigr){\,}$.  When the limit
$R\rightarrow\infty$ is eventually taken, one expects that the
contribution from $C_R$ to the total action should vanish; this
requires that ${\mid}f_{\ell m}(k){\mid}$ should decay at least as
rapidly as ${\mid}k{\mid}^{-2}$, as
${\mid}k{\mid}\rightarrow\infty{\,}$.  In fact, on dimensional
grounds, one expects that
$${\mid}f_{\ell m}(k){\mid}{\;}{\,}\sim{\;}{\,}{\mid}k{\mid}^{-3}\eqno(4.6)$$
as ${\,}{\mid}k{\mid}\rightarrow\infty{\,}$.  To see this, rewrite the 
radial equation (2.26) in terms of the operator
$${\cal L}_{\ell}{\;}{\,} 
={\;}{\,}e^{(b-a)/2}{\;}
{{d}\over{dr}}\biggl(e^{(b-a)/2}{\;}{{d}\over{dr}}({\;}{\,})\biggr) 
-V_{\ell}(r){\quad},\eqno(4.7)$$
which is self-adjoint with respect to the inner product (3.5).  Then
note that Eq.(3.13) can be rewritten as
$$a_{k\ell m}{\,}+{\,}a_{-k\ell m}{\;}{\,} 
={\;}{\,}{{-1}\over{2\pi k^{2}{\,}{\mid}z_{k\ell}{\mid}^{2}}}{\;}
\int^{R_{\infty}}_{0}{\,}dr{\;}{\,}e^{(a-b)/2}{\;}\xi_{k\ell}(r){\;}{\,} 
{\cal L}_{\ell}\psi_{\ell m}(r){\quad}.\eqno(4.8)$$
\noindent
We have used the boundary condition (3.1) and assumed that 
$\psi_{\ell m}(r)$ dies out at large $r{\,}$.  The form (4.8) is just 
an expression of the self-adjointness of the radial equation.  Now consider the
dimensions of the quantities involved -- treated explicitly in Sec.5
below.  In particular, $\psi_{\ell m}(r)$ has dimensions of length and
${\mid}z_{k\ell}{\mid}^2$ is dimensionless.  In the limit 
$R_{\infty}\rightarrow\infty{\,}$, and for large $k$ (so taking a WKB 
approximation for the radial functions), the integral in Eq.(4.8) can 
only involve the dimensionless frequency $2Mk{\,}$, where $M$ is the 
total mass (true ADM mass) of the space-time. This gives the desired 
behaviour (4.6) at large ${\mid}k{\mid}{\,}$.

The contour $C_{\epsilon}$ gives a purely imaginary contribution to the
total Lorentzian action; also (below) the curve $C_{\alpha}$ gives a
complex contribution.  We shall interpret the quantity  
$\exp\bigl[-2{{\,}\rm Im}(S)\bigr]{\,}$, up to normalisation, as
describing the conditional probability density over the final boundary 
data.  To compute $J_{\epsilon}{\,}$, we assume that $f_{\ell m}(k)$ 
is analytic in a neighbourhood of 
${\,}k=\sigma_n{\,}$, where
$$\sigma_{n}{\;}{\,}={\;}{\,}{{n\pi}\over{{\mid}T{\mid}}}{\quad},\eqno(4.9)$$
for $n=1,2,\ldots{\;}$.  Then
$$\eqalign{J_{\epsilon}{\;}{\,}&
={\;}{\,}-{\;}\lim_{\epsilon\rightarrow 0}{\;}\sum_{\ell m}{\;}
\int_{C_\epsilon}{\,}dk{\;}{\,}f_{\ell m}(k){\;}
\cot\bigl(k{\mid}T{\mid}\bigr)\cr
&={\;}{\,}{{i\pi}\over{{\mid}T{\mid}}}{\;}\sum_{\ell m}{\;}
\sum^{\infty}_{n=1}{\;}{\,}f_{\ell m}(\sigma_n){\quad}.\cr}\eqno(4.10)$$
\indent
For the curve $C_\alpha{\,}$, one has 
$$J_{\alpha}{\;}{\,} 
={\;}{\,}-{\;}\sum_{\ell m}{\;}\int^{R}_{0}{\;}d{\mid}k{\mid}{\;}{\,}
e^{-i\alpha}{\;}f_{\ell m}\Bigl({\mid}k{\mid}{\,},\alpha\Bigr){\;}
\cot\Bigl({\mid}k{\mid}{\,}e^{-i\alpha}{\,}{\mid}T{\mid}\Bigr){\quad}.
\eqno(4.11)$$ 
\noindent
We shall need the properties [14]
$$\cot(x){\;}{\,} 
={\;}{\,}\sum^{\infty}_{n=-\infty}{\;}{{1}\over{(x-n\pi)}}
\eqno(4.12)$$
and
$${{1}\over{(x-a\pm i\epsilon)}}{\;}{\,} 
={\;}{\,}{\rm P.P.}{\;}{1\over(x-a)}{\;}\mp{\;}i\pi{\,}\delta(x-a)
{\quad},\eqno(4.13)$$
\noindent
where ${\rm P.P.}$ denotes the principal part.  Assuming that 
$f_{\ell m}(k)$ is regular along $C_\alpha{\,}$, one has, for small 
$\alpha{\,}$:
$$J^{(1)}_{\alpha}{\;}{\,} 
={\;}{\,}\lim_{\alpha\rightarrow 0_{+}}{\,}(1-i\alpha){\;}
\sum_{\ell m}{\;}\sum^{\infty}_{n=-\infty}{\;}\int^{R}_{0}{\,} 
d{\mid}k{\mid}{\;}{\,}{{f_{\ell m}\bigl({\mid}k{\mid}{\,},\alpha\bigr)}
\over{\bigl({\mid}kT{\mid}-n\pi-i\alpha\bigr)}}\eqno(4.14)$$
\noindent
In the further limit $R\rightarrow\infty{\,}$, this gives
$$J_{\alpha}{\;}{\,} 
={\;}{\,}{\rm P.V.}{\;}+{\,}{{i\pi}\over{{\mid}T{\mid}}}{\;} 
\sum_{\ell m}{\;}\sum^{\infty}_{n=1}{\;}f_{\ell m}(\sigma_n){\quad},
\eqno(4.15)$$
\noindent
where ${\rm P.V.}$ denotes the principal-value part of the integral.

Using Eqs.(4.4,5,10,15), the classical action for massless
scalar-field perturbations, in the case that 
${\,}T{\,}={\,}{\mid}T{\mid}\exp(-i\delta){\,}$ is very slightly
complex, is
$$\eqalign{S^{(2)}_{\rm class}\Bigl[\{a_{k\ell m}\};
{\mid}T{\mid}\Bigr]{\;}{\,}&
={\;}{\,}{\rm real{\,}part}{\;} 
+{\,}{{2i\pi}\over{{\mid}T{\mid}}}{\;}\sum^{\infty}_{\ell=0}{\;}
\sum^{\ell}_{m=-\ell}{\;}\sum^{\infty}_{n=1}{\;}{\,}
f_{\ell m}(\sigma_{n})\cr
&={\;}{\,}{\rm real{\,}part}{\;}+{\,}{{2i\pi^{2}}\over{{\mid}T{\mid}}}{\;}
\sum_{\ell mn}{\;}{\,}\sigma_{n}{\,}{\mid}z_{n\ell}{\mid}^{2}{\;}
{\mid}a_{n\ell m}{\,}+{\,}a_{-n\ell m}{\mid}^{2}{\quad}.\cr}\eqno(4.16)$$
\noindent
The real part of $S^{(2)}_{\rm class}$ is, of course, also calculable
from the equations above.  It contains the principal-value term and
the real part of Eq.(4.10). The main, semi-classical contribution to the
quantum amplitude is then 
${\,}\exp\Bigl(iS^{(2)}_{\rm class}\bigl[\{a_{k\ell m}\};
{\mid}T{\mid}\bigr]\Bigr)$.  The probability distribution for final
configurations involves only 
${\rm Im}\bigl(S^{(2)}_{\rm class}\bigr)$;
the more probable configurations will have 
${\,}S^{(2)}_{\rm class}{\,}$ 
lying only infinitesimally in the upper half-plane.  Whether probable 
or not, those final configurations  $\{a_{k\ell m}\}$ which contribute to the
probability distribution must yield finite expressions in the infinite
sums over $n\ell$ in Eq.(4.16).  There will be a corresponding
restriction when the data  are instead described in terms of the
spatial configurations $\{\psi_{\ell m}(r)\}$. Also, as can be seen in 
[17], the complex quantities 
$z_{n\ell}{\,}(a_{n\ell m}{\,}+{\,}a_{-n\ell m})$ 
appearing in Eq.(4.16) are related to Bogoliubov transformations
between initial and final states, thus providing a further
characterisation of the finiteness of 
${\rm Im}\bigl(S^{(2)}_{\rm class}\bigr)$ in Eq.(4.16).

With regard to the sum over $\ell$ in Eq.(4.16), one imagines that a
cut-off $\ell_{{\rm max}}$ can be provided by the radial equation (2.26).
In the region where $\bigl(V_{\ell}(r)-k^2\bigr)>0{\,}$, one has
exponentially growing radial functions, whereas for 
$\bigl(V_{\ell}(r)-k^{2}\bigr)<0{\,}$, one has oscillatory radial 
functions. One defines $\ell_{\rm max}$ by
$\bigl(V_{\ell{\rm max}}(r)-k^{2}\bigr)=0{\,}$, and restricts attention
mainly to oscillatory solutions.

When one has both initial and final non-zero Dirichlet data labelled by
'coordinates' $\{a^{(I)}_{k\ell m}\}$ and $\{a^{(F)}_{k\ell m}\}$, the
perturbative classical scalar action $S^{(2)}_{\rm class}$ includes
separate terms of the form (4.16) for the initial and final data.  But
$S^{(2)}_{\rm class}$ also includes a cross-term between
$a^{(I)}_{k\ell m}$ and $a^{(F)}_{k\ell m}{\,}$, which represents the
correlation or mixing between the initial and final data.  The total
action will naturally be symmetric in $a^{(I)}_{k\ell m}$ and
$a^{(F)}_{k\ell m}{\,}$, and the coefficients $z_{n\ell}$ will be the same
(they are time-independent) up to a phase.  For large ${\mid}T{\mid}{\,}$,
the cross-term becomes negligible, and one has two independent
contributions to the classical action, one being a functional of
$\{a^{(I)}_{k\ell m}\}$, the other of $\{a^{(F)}_{k\ell m}\}$.

\end{section}

\begin{section}{Dimensional analysis}

In Sec.4 above, we made use of a dimensional argument, in order to
obtain the estimate (4.6) of the rate of fall-off of 
${\mid}f_{\ell m}(k){\mid}$ for large ${\mid}k{\mid}{\,}$. In the 
present Section, we give a more thorough treatment of the 
dimensionality of the main quantities appearing in the preceding 
paper I, this Paper II and subsequent work.

The classical perturbative Einstein equations are schematically of the
form $L^{-2}\sim\bigl(\partial\phi^{(1)}\bigr)^{2}$, where $L$ is 
a length-scale characteristic of the gravitational field.   Hence, the
massless-scalar perturbations $\phi^{(1)}$ are dimensionless.
Similarly, one finds that the metric perturbations $h^{(1)}_{\mu\nu}$
are dimensionless.  From this, one deduces that the functions
$\xi_{k\ell m}(t,r)$ of Eq.(2.16), appearing in the adiabatic
approximation, are also dimensionless, as are the complex coefficients
$z_{k\ell}$ of Eq.(3.3).

In describing dimensionful quantities, let us denote by $M$ the ADM
mass of the 'space-time'.  We recall again the distinction which it
was necessary to draw in Sec.I.5 and in [3] between the naive mass, computed
from the fall-off of the spatial metric $g_{ij}$ of a hypersurface
badly embedded near spatial infinity, and the true ADM mass. Here, $M$
denotes the true ADM mass, as given by the standard definition for a
hypersurface which approaches spatial infinity in the familiar way
[10].  Suppose that the time-separation $T$ at spatial infinity obeys
${\mid}T{\mid}\gg 2M{\,}$, as will normally be the case; further, let the
radius $R_{\infty}$ tend to infinity.  Then, in our treatment of
massless perturbations, $M$ can be regarded as setting a reference
length-scale.

We now write
$$x{\;}{\,}={\;}{\,}2Mk\eqno(5.1)$$
for the dimensionless frequency.  Further, recall that the coordinates
$\{a_{k\ell m}\}$ of Eq.(2.18) for the perturbative scalar field on the
final surface $\Sigma_F$ are defined irrespective of $T$.  From
Eq.(2.18), it follows that each $a_{k\ell m}$ scales as $M^2$.  That is,
one can write
$$a_{k\ell m}{\;}{\,}={\;}{\,}M^{2}{\;}y_{\ell m}(x){\quad},\eqno(5.2)$$
for some (smooth) dimensionless complex functions $y_{\ell m}(x)$.
This property can also be derived from Eq.(3.13). By following these
arguments, one can write the imaginary part of the total classical
action as
$${\rm Im}\biggl(S^{(2)}_{\rm class}
\Bigl[\{y_{\ell m}\};{\mid}T{\mid}\Bigr]\biggr){\;}{\,} 
={\;}{\,}\biggl({\pi\over2}\biggl){\;}M^{2}{\;}
\sum_{n\ell m}{\;}x_{n}{\;}(\Delta x_{n}){\;}
{\mid}z_{n\ell}{\mid}^{2}{\;}{\mid}y_{n\ell m}{\mid}^{2}{\quad},\eqno(5.3)$$
\noindent
where
$$\eqalignno{x_{n}&{\;}{\,}={\;}{\,}2M\sigma_{n}{\quad},
{\qquad}{\quad}\sigma_{n}{\;}{\,}
={\;}{\,}{{n\pi}\over{{\mid}T{\mid}}}{\quad},&(5.4)\cr  
\Delta x_{n}{\;}{\,}&={\;}{\,}{{2\pi M}\over{{\mid}T{\mid}}}{\quad}.
&(5.5)\cr}$$
\indent
Clearly, some understanding of the numerical magnitude of the sum in
Eq.(4.16) is necessary, if 
$\exp\bigl[-2{\,}{\rm Im}(S^{(2)}_{\rm class})\bigr]$
is to be interpreted as a conditional probability density. 
(Indeed, the requirement that these probabilities sum to $1$ is part 
of the requirement of unitarity!)  In this lowest-order perturbative
approximation, the probability density above for scalar fluctuations
is Gaussian; the same holds in the present Einstein/massless-scalar
case for the distribution of gravitational-wave fluctuations on the
final surface $\Sigma_F{\,}$; this is treated in more detail in [18].
The Gaussian property is a perturbative aspect of a more general 
property -- arising in the Positive Action conjecture [19,20]. 
For simplicity of exposition, let us consider this for the case
of Riemannian 4-metrics, although we expect some version to hold in
the more general complex case. Suppose that, as in the 'Euclidean'
path integral, we take all smooth 4-metrics $g_{\mu\nu}{\,}$, 
scalar fields $\phi,\ldots{\;}$ which agree with Dirichlet boundary data
$h_{ij},\;\phi,\ldots{\;}$ prescribed on a compact boundary $\partial V$,
bounding a compact manifold-with-boundary $\bar V$.  The Riemannian action
functional $I[g_{\mu\nu};{\,}\phi;\ldots{\;}]$ may be seen to take
arbitrarily negative values, when one applies suitable high-frequency
conformal transformations 
${\,}g_{\mu\nu}\rightarrow\Omega^{2}g_{\mu\nu}{\,}$ to the metric [19]. 
But, at least for Einstein gravity with the simplest topology, 
the classical action $I_{\rm class}$ is a non-negative functional 
of the boundary data [20].  Similar positivity properties can be 
investigated in  the case of Einstein gravity coupled to matter.

This dimensional analysis sets a natural normalisation for the (equal) 
quantities on the left-hand sides of Eqs.(3.10,11).  Using Eq.(5.2), 
one has 
$$\eqalign{2{\;}\sum_{\ell m}{\;}\int^{\infty}_{0}{\,}dk{\;}{\,}
{\mid}z_{k\ell}{\mid}^{2}{\;}{\mid}a_{k\ell m}+a_{-k\ell m}{\mid}^{2}
{\;}{\,}&={\;}{\,}M^{3}{\;}\sum_{\ell m}{\;}
\int^{\infty}_{0}{\,}dx{\;}{\,}{\mid}z_{\ell}(x){\mid}^{2}{\;}
{\mid}y_{\ell m}(x){\mid}^{2}\cr
&\propto{\;}{\,}t_{0}{\quad}.\cr}\eqno(5.6)$$
\noindent
The last line applies to semi-classical collapse and subsequent
evaporation of a non-rotating black hole, where 
${\,}t_{0}{\;}\propto{\;}M^{3}{\,}$ is the time taken for complete 
evaporation [15].

\end{section}

\begin{section}{Conclusion}

In this paper, we have derived the quantum amplitude [through
Eq.(4.16)] for a spherically-symmetric configuration $(h_{ij},\phi)_I$ 
on the initial surface $\Sigma_I$ to become a given configuration 
$(h_{ij},\phi)_F$ on the final surface $\Sigma_F{\,}$, with Lorentzian 
time-interval $T$ at spatial infinity, provided that the final 
3-dimensional metric $h_{ijF}$ is also spherically symmetric.  
In the amplitude, which is of the form
$({\rm const.}){\times}\exp\bigl(iS^{(2)}_{\rm class}\bigr)$, the action
$S^{(2)}_{\rm class}$ depends approximately quadratically on the 
(non-spherical) perturbative part of the final data $\phi_F{\,}$.  
Further, $S^{(2)}_{\rm class}$ has both a real part and an imaginary part.
The imaginary part leads to a Gaussian probability density 
${\,}{\mid}\Psi{\mid}^{2}{\,}\propto{\,}
\exp\bigl(-2{\,}{\rm Im}\bigr(S^{(2)}_{\rm class}\bigr)\bigr)$, while the
real part gives rapid oscillations through the phase of the quantum
amplitude or wave function $\Psi$.  Corresponding quantum amplitudes
for spin-1 and spin-2 fields have been calculated [18], while the 
fermionic massless spin-$1\over 2$ case is summarised in [21].

The familiar description of black-hole evaporation in terms of
Bogoliubov coefficients occurs in many works [22-24].  Following the 
'Complex Approach' and the 'Spin-0 Amplitude' for black-hole
evaporation, we next make a
connection between the description and calculation of quantum
amplitudes in the present paper and the alternative description in
terms of Bogoliubov coefficients [17].  Further, we treat the 
approximation of the radiative part of the space-time by the Vaidya 
metric [12], related to the description of adiabatic scalar-field 
solutions in Sec.2 of the present paper.  Yet another description of 
quantum amplitudes will later be provided, relating Eq.(4.16) to
coherent and squeezed states; this gives a more general conceptual 
framework with which we have already examined the amplitudes of the 
present paper and the higher-spin amplitudes [18,21].
\end{section}

\begin{section}*{References}
\everypar{\hangindent\parindent}

\noindent[1]  A.N.St.J.Farley and P.D.D'Eath, 'Quantum Amplitudes in
Black-Hole Evaporation: I. Complex Approach', submitted for
publication, 2005.

\noindent[2] A.N.St.J.Farley, 'Quantum Amplitudes in Black-Hole 
Evaporation', Cambridge Ph.D. dissertation, approved 2002 (unpublished);
A.N.St.J.Farley and P.D.D'Eath, Phys Lett. B, {\bf 601}, 184 (2004).

\noindent[3] M.K.Parikh and F.Wilczek, Phys. Rev. Lett. {\bf 85}, 5042
(2000); M.Parikh, Gen. Relativ. Gravit. {\bf 36}, 2419 (2004).

\noindent[4] P.R.Garabedian, {\it Partial Differential Equations}, 
(Wiley, New York) (1964).

\noindent[5] P.D.D'Eath, {\it Supersymmetric Quantum Cosmology}, 
(Cambridge University Press, Cambridge) (1996).

\noindent[6] R.P.Feynman and A.R.Hibbs, {\it Quantum Mechanics and Path
Integrals}, (McGraw-Hill, New York) (1965).

\noindent[7] W.McLean, {\it Strongly Elliptic Systems and Boundary
Integral Equations}, (Cambridge University Press, Cambridge) (2000); 
O.Reula, 'A configuration space for quantum gravity and solutions to
the Euclidean Einstein equations in a slab region',
Max-Planck-Institut f\"ur Astrophysik, {\bf MPA}, 275 (1987).

\noindent[8] P.D.D'Eath, 'Loop amplitudes in supergravity by canonical
quantization', in {\it Fundamental Problems in Classical, Quantum and
String Gravity}, ed. N.S\'anchez (Observatoire de Paris) 166 (1999), 
hep-th/9807028.

\noindent[9] P.D.D'Eath, 'What local supersymmetry can do for quantum
cosmology', in {\it The Future of Theoretical Physics and Cosmology},
eds. G.W.Gibbons, E.P.S.Shellard and S.J.Rankin (Cambridge University
Press, Cambridge) 693 (2003).

\noindent[10] C.W.Misner, K.S.Thorne and J.A.Wheeler, {\it Gravitation},
(Freeman, San Francisco) (1973).

\noindent[11] J.A.H.Futterman, F.A.Handler and R.A.Matzner,   
{\it Scattering from Black Holes} (Cambridge University Press,
Cambridge) (1988).

\noindent[12] P.C.Vaidya,  Proc. Indian Acad. Sci. {\bf A33}, 264
(1951); R.W.Lindquist, R.A.Schwartz and C.W.Misner, Phys. Rev. 
{\bf 137}, 1364 (1965); A.N.St.J.Farley and P.D.D'Eath, 'Vaidya
Space-Time and Black-Hole Evaporation', submitted for publication (2005).

\noindent[13] R.M.Wald, Phys. Rev. D {\bf 13}, 3176 (1976).

\noindent[14] M.Abramowitz and I.A.Stegun, {\it Handbook of Mathematical
Functions}, (Dover, New York) (1964).

\noindent[15] D.N.Page and S.W.Hawking, Astrophys.J. {\bf 206}, 1 (1976).

\noindent[16] T.Regge and J.A.Wheeler, Phys. Rev. {\bf 108}, 1063 (1957).

\noindent[17]  A.N.St.J.Farley and P.D.D'Eath, Phys. Lett B 
{\bf 613}, 181 (2005).

\noindent[18] A.N.St.J.Farley and P.D.D'Eath,
Class. Quantum. Grav. {\bf 22}, 2765 (2005).

\noindent[19] G.W.Gibbons, S.W.Hawking and M.J.Perry, Nucl. Phys. {\bf B138},
141 (1978).

\noindent[20] R.Schoen and S.-T.Yau, Phys. Rev. Lett. {\bf 42}, 547 (1979).

\noindent[21] A.N.St.J.Farley and P.D.D'Eath, Class.Quantum Grav.
 {bf 22}, 3001 (2005).

\noindent[22] S.W.Hawking,  Commun. Math. Phys. {\bf 43}, 199 (1975).

\noindent[23] N.D.Birrell and P.C.W.Davies, {\it Quantum fields in curved
space}, (Cambridge University Press, Cambridge) (1982).

\noindent[24] V.P.Frolov and I.D.Novikov, {\it Black Hole Physics}, (Kluwer
Academic, Dordrecht) (1998).

\end{section}

\end{document}